# Ubiquitous Ideal Spin-Orbit Coupling in a Screw Dislocation in Semiconductors

Lin Hu[1*], Huaqing Huang[2*], Zhengfei Wang[3], W. Jiang[2], Xiaojuan Ni[2], Yinong Zhou[2], V. Zielasek[4η], M.G. Lagally[4], Bing Huang[1#], and Feng Liu[2,5#]

[1]*Beijing Computational Science Research Center, Beijing 100193, China*

[2]*Department of Materials Science and Engineering, University of Utah, Salt Lake City, UT 84112, USA*

[3]*Hefei National Laboratory for Physical Sciences at the Microscale, University of Science and Technology of China, Hefei, Anhui 230026, China*

[4]*Department of Materials Science and Engineering, University of Wisconsin, Madison, WI 53706, USA*

[5]*Collaborative Innovation Center of Quantum Matter, Beijing 100084, China*

*These authors contributed equally to this work.

[η]Current address: *Institute of Applied and Physical Chemistry, University of Bremen, Bremen, D-28359, Germany*

[#]Correspondence to: F.L. (fliu@eng.utah.edu) or B.H. (bing.huang@csrc.ac.cn)

Abstract

We theoretically demonstrate that screw dislocation (SD), a 1D topological defect widely present in semiconductors, exhibits ubiquitously a new form of spin-orbit coupling (SOC) effect. Differing from the widely known conventional 2D Rashba-Dresselhaus (RD) SOC effect that typically exists at surfaces/interfaces, the deep-level nature of SD-SOC states in semiconductors readily makes it an ideal SOC. Remarkably, the spin texture of 1D SD-SOC, pertaining to the inherent symmetry of SD, exhibits a significantly higher degree of spin coherency than the 2D RD-SOC. Moreover, the 1D SD-SOC can be tuned by ionicity in compound semiconductors to ideally suppress spin relaxation, as demonstrated by comparative first-principles calculations of SDs in Si/Ge, GaAs, and SiC. Our findings therefore open a new door to manipulating spin transport in semiconductors by taking advantage of an otherwise detrimental topological defect.

Spintronics offers a promising paradigm shift for future information and energy technologies by processing the electron spin instead of charge degree of freedom, thereby essentially avoiding heat dissipation. In a crystalline solid, the motion of an electron is inevitably coupled with its spin orientation through the spin-orbit coupling (SOC) effect. Therefore, discovering new forms of the SOC effect that provide more effective means to manipulate spin transport properties is not only of fundamental interest but also critical to the development of spintronics devices.

In a crystal that lacks inversion symmetry, the electronic energy bands will lift spin degeneracy as a manifestation of the Rashba [1] and/or Dresselhaus [2] SOC effect, which provides a useful physical mechanism to manipulate spin transport by an electrical field. The Rashba-Dresselhaus SOC (RD-SOC) was initially demonstrated in semiconductor heterostructures, e.g., the asymmetric InGaAs/InAlAs quantum well [3], and later extended to noble-metal [4-6] or heavy-metal surfaces [7-9]. However, to exploit RD-SOC in spintronics devices, two challenging requirements have to be met. First, the spin-polarized SOC states are often not ideal, in the sense that they overlap with other, spin-unpolarized, host states around the Fermi energy, as exemplified in the strong hybridization between the Rashba surface states and the substrate "bulk" states [4-9]. The ideal Rashba state has been realized only in the Te-terminated surface of BiTeX (X=I, Br and Cl) [10] and has been proposed in strained interfaces [11]. Second, the spin texture of the original form of Rashba or Dresselhaus SOC suffers from spin randomization (relaxation) caused by momentum-changing electron scattering, and combining these two forms of SOC with equal magnitude is needed to produce an 'ideal' spin texture to suppress spin relaxation [12].

In this Letter we demonstrate a new form of SOC, a 1D SOC effect that ubiquitously exists in screw dislocations (SD) in semiconductors, to meet both these ideal conditions. Dislocations exist inevitably in all the semiconductors. They fall into three categories, i.e., edge-, mixed- and screw-type, defined by their Burgers vector *b*, which depicts the direction and amount of slip associated with each type of dislocation. In electronic and optoelectronic devices, dislocations are undesirable, because they degrade the device functionality. For example, in transistors dislocations cause current leakage, while in light-emitting diodes and solar cells dislocations induce in-gap deep defect levels acting as deleterious carrier recombination centers [13-15]. As such, much research effort in the past has been devoted to alleviating active dislocations in semiconductors. Surprisingly, here we defy the conventional wisdom by turning these ordinarily harmful dislocations into something useful.

We describe a ubiquitous SOC effect associated with all SDs in semiconductors that we have discovered. We realized that a SD, a 1D spiral-type line defect in a crystal, assumes always a rotational plus a fractional translational symmetry, which locally breaks inversion symmetry and generates an effective electrical field pertaining to the inherent symmetry of SD. This leads to an *ideal* form of SOC as the 1D SD-SOC states reside deep in the band gap of semiconductors, and are thus completely isolated from the host semiconductor bands. Furthermore, the spin texture of the 1D SD-SOC exhibits intrinsically a higher degree of spin coherency than the conventional 2D RD-SOC, as the spin orientations are constrained to vary within a much narrower range of $(0, \pi/2)$ rather than $(0, 2\pi)$. The spin texture is also tunable by the ionicity of a compound semiconductor, to achieve an *ideal* persistent momentum-independent orientation to suppress spin relaxation, as demonstrated by comparative first-principles calculations of several representative semiconductors (Si/Ge, GaAs, and SiC). Therefore, this new form of 1D SD-SOC exhibits two *ideal* features that are not only fundamentally interesting but also highly desirable for spintronics device applications.

We first introduce the key features of 1D SD-SOC effect and highlight its main difference from conventional 2D RD-SOC effect, as illustrated in Fig. 1. In general, the SOC effect in a crystal can be described by the Hamiltonian $\mathrm{H_{SOC}} = \frac{\alpha}{\hbar}(\boldsymbol{E} \times \boldsymbol{p}) \cdot \boldsymbol{\sigma}$ and $\boldsymbol{E}=\nabla \mathrm{V}$, where $\alpha$ is the material dependent SOC constant, $h$ is Plank constant, $\boldsymbol{E}$ is the effective field induced by gradient of potential (V) in an inversion-asymmetry system, $\boldsymbol{p}$ is the momentum, and $\boldsymbol{\sigma} = (\sigma_x, \sigma_y, \sigma_z)$ are the Pauli matrices. For a system with $C_{2v}$ rotational symmetry, a 2D RD-SOC Hamiltonian is derived as [16]:

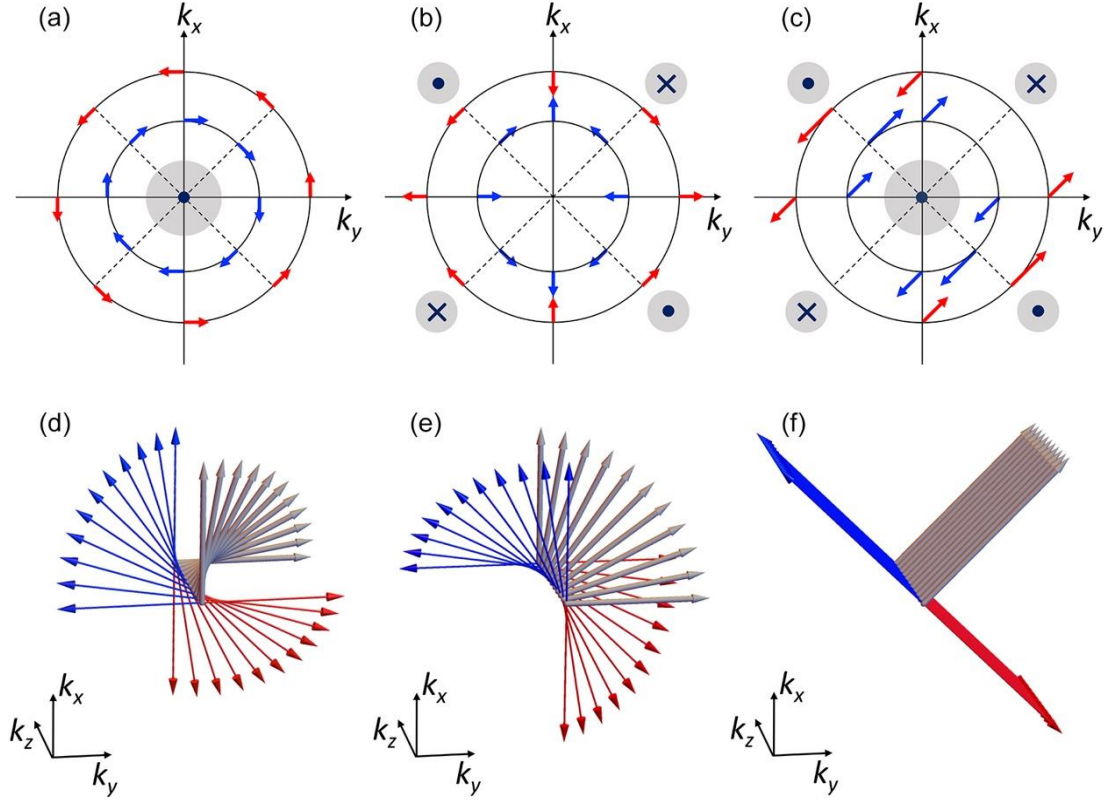

**Fig. 1.** (a) The orientations of the effective electrical field (light grey shaded circle) and spins (red and blue arrows) for the conventional Rashba SOC effect at surfaces/interfaces. (b) Same as (a) for the linear Dresselhaus effect in an asymmetry QW or a strained zinc-blende film. (c) Combined effect of (a) and (b). (d) The orientations of the effective electrical field (grey arrows) and spins (red and blue arrows) for the 1D SD-SOC effect as found in Ge. (e) Same as (d) as found in GaAs. (f) Combined effect of (d) and (e) as found in SiC.

$$H_{RD}^{2D} = \lambda_R \left(k_y \sigma_x - k_x \sigma_y\right) + \lambda_D \left(k_x \sigma_x - k_y \sigma_y\right), \qquad (1)$$

where $k_x$ and $k_y$ are the reciprocal-space wave vectors, and $\lambda_R$ and $\lambda_D$ are the Rashba and Dresselhaus SOC strengths, respectively. The first Rashba term usually arises at a surface or interface [3,10], where there exists a $k$-independent effective electric field in the direction perpendicular to the surface/interface, i.e., $\boldsymbol{E} = e\hat{\mathbf{z}}$. Then the spin-polarized bands adopt a 2D spin texture in the ($k_x$, $k_y$) plane as shown in Fig. 1(a). The second linear Dresselhaus term arises in an asymmetric quantum well or a strained zinc-blende film [17], where there is no macroscopic "net" electrical field but effective *k*-dependent local fields, $\boldsymbol{E}(-k_x, k_y) = \boldsymbol{E}(k_x, -k_y) = e\hat{z}$ and $\boldsymbol{E}(k_x, k_y) = \boldsymbol{E}(-k_x, -k_y) = -e\hat{z}$, and a corresponding 2D spin texture in the ($k_x$, $k_y$) plane as shown in Fig. 1(b). In a system with both these two terms present with equal strength, the spin texture has a *k*-independent fixed spin orientation as shown in Fig. 1(c).

In the 1D SD system, the potential $V$ and hence the effective field $\boldsymbol{E}$ also share the same screw symmetry as the spiral-type structure of the dislocation. Taking the general 2-fold screw rotation symmetry along *c*-axis $\hat{S} = \{C_2 | \frac{c}{2}\}$, the potential becomes $V = V(\cos\vartheta, \sin\vartheta)$, where $\vartheta = \frac{2\pi}{c/2} \cdot z = 2 \cdot G_z \cdot z$. After some algebra (see Supplementary Information for details [18]), a general 1D-SD SOC Hamiltonian can be expressed as

$$H_{SD}^{1D} = \lambda_e k_z \left(\cos\frac{k_z}{2}\sigma_y - \sin\frac{k_z}{2}\sigma_x\right) + \lambda_c k_z \left(\cos\frac{k_z}{2}\sigma_x - \sin\frac{k_z}{2}\sigma_y\right), \quad (2)$$

where $k_z$ is the reciprocal-space wave vector in units of $\pi/c$. $\lambda_e$ and $\lambda_c$ are the SOC strengths for the first and second term, respectively. These two terms are found to correspond to the SD-SOC Hamiltonian in elemental semiconductors and compound semiconductors with high ionicity, respectively, as we show later with first-principles materials calculations and tight-binding (TB) model calculations fitting the first-principles results. A key difference between Eq. (2) and Eq. (1) is a sinusoidal dependence of the effective field and hence spin texture on $k_z$, which results in a spin rotation in the ($k_x$, $k_y$) plane with a period of $\pi$. Because the SD has a $C_2$ rotation plus translation symmetry, the effective field and hence the spin texture must be subject to this symmetry. Solving Eq. (2) we obtain the spin texture of the 1D SD-SOC [18]. The first term gives rise to the following spin orientation as a function of $k_z$:

$$\vec{S}_{1,2} = \mp \sin\frac{k_z}{2}\hat{k}_x \pm \cos\frac{k_z}{2}\hat{k}_y, \quad (3a)$$

$$\theta_1 = \frac{\pi}{2} + \frac{k_z}{2}, \quad \theta_2 = -\frac{\pi}{2} + \frac{k_z}{2}, \quad (3b)$$

where subscripts 1 and 2 represent spin up and down, respectively. θ is the angle between the spin and $k_x$ axis. Similarly, from the second term in Eq. (2),

$$\vec{S}_{1,2} = \pm\cos\frac{k_z}{2}\hat{k}_x \mp \sin\frac{k_z}{2}\hat{k}_y, \quad (4a)$$

$$\theta_1 = \pi - \frac{k_z}{2}, \quad \theta_2 = -\frac{k_z}{2}. \quad (4b)$$

When both terms in Eq. (2) are present with equal strength, one has

$$\theta_1 = \frac{3\pi}{4}, \quad \theta_2 = -\frac{\pi}{4}. \quad (5)$$

These results lead to spin textures with spins rotating within one of the four quadrants, as shown in Fig. 1(d) and 1(e) for the first and second terms in Eq. (2), respectively. The spin texture of these two terms differs by a phase of π/2 (see Fig. 1(d) versus Fig. 1(e)). A very interesting case is when both terms are present with equal strength, then the effectively field and hence the spin orientation becomes fixed along the diagonal direction of a quadrant independent of $k_z$, as shown in Fig. 1(f).

There are significant implications of different spin textures, as shown in Fig. 1, on spin relaxation and coherence time. In a solid, due to the spin-momentum locking property, spin will rotate when electrons are scattered with a sudden change of momentum, leading to spin relaxation and a shortened spin coherence time. For the conventional 2D RD-SOC as shown in Fig. 1(a) and 1(b), spins can rotate in four quadrants in the ($k_x$, $k_y$) plane with an angle changing from 0 to 2π. In contrast, for the newly discovered 1D SD-SOC as shown in Fig. 1(d) and 1(f), spins can only rotate in one of the four quadrants with an angle changing from 0 to π/2. Consequently, the 1D SD-SOC will exhibit a significant higher degree of spin coherency because the spins are constrained to vary within a much narrower range of angles. Furthermore, it is known that for the 2D RD-SOC, to suppress spin relaxation an ideal spin texture as shown in Fig. 1(c) can be engineered by including both the first and second terms in Eq. (1) with equal strength, such as in a III-V heterostructure [19]. Similarly, this ideal spin texture can also be achieved with the 1D SD-SOC as shown in Fig. 1(e), albeit be available intrinsically in a single material rather than at a heterostructure.

In the following, to confirm the above theoretical analysis, we will characterize and quantify the 1D SD SOC effects in several representative semiconductors with different levels of ionicity, including Ge (Si), GaAs, and SiC, using first-principles density-functional-theory and TB model Hamiltonian calculations [18]. Furthermore, to support our theoretical studies, we demonstrate also an experimental approach to show the feasibility of controlled formation of SDs in Si [18].

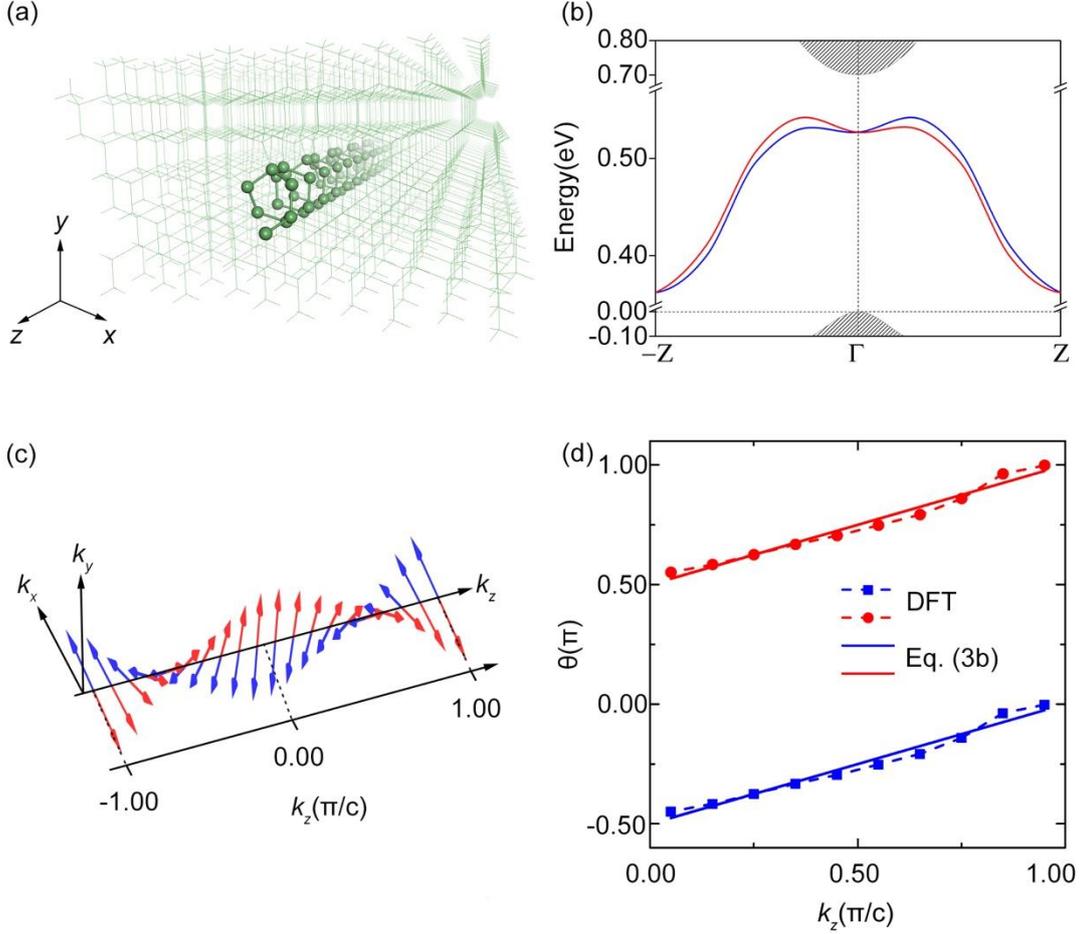

**Fig. 2.** (a) The atomic structure of a SD in bulk Ge. (b) Band structure of a SD in Ge, with SOC effect. The Fermi level is set to zero. The red and blue lines represent SD-SOC bands with different spin projections. (c) The spin textures of SD-SOC bands obtained from DFT calculations. The red and blue arrows shows orientations of two spin projections. (d) Data points (dots) show θ, the angle between spin and the $k_x$ axis, as a function of $k_z$ obtained from DFT calculations. The lines are drawn according to Eq. (3b).

We first present the results of SD in Ge (Fig. 2). For the case of Ge (Si), we consider the pure SDs whose <110> **b** lies in the {111} slip plane [20] (see Fig. S1 [18]). In our model, the SDs of different separations are constructed in orthorhombic supercells of the same size, containing 480 atoms and 66.5Å × 18.8Å in dimension in the (110) plane perpendicular to the **b**. It is worth noting that to satisfy the periodic boundary condition, two SDs (a dislocation dipole) with opposite Burgers vectors are necessary to be included in one supercell [21]. They are set sufficiently far apart to avoid annihilation in the periodic supercell.

Si and Ge have similar geometric structures, and the SDs in both systems are almost the same. Because of the larger SOC effect in Ge, we will show the results of Ge here while leaving the results for Si in the Fig. S2 [18]. The SDs are known to induce deep in-gap defect levels, as shown in Fig. 2(a). Excluding SOC in the calculation, the two bands are spin degenerate, with charge distributions that are solely determined by the Ge atoms in the dislocation core (see Fig. S3 [18]). Time reversal symmetry (TRS) ensures $E_\sigma(k) = E_{\bar{\sigma}}(-k)$ for Kramers doublets with opposite momenta and orthogonal spins. Including SOC, the spin degeneracy of these two defect bands is lifted, as shown in Fig. 2(b). In Fig. 2(c), we show the spin texture $\vec{p}(\vec{k})$ of the SOC-lifted states induced by the SD in Ge. Blue and red colors represent the two bands with opposite spin orientations. At the same *k*, the directions of spins of lower and higher energy bands are opposite. Within the same band, the spin texture satisfies the condition, $\vec{p}(\vec{k}) = -\vec{p}(-\vec{k})$, ensured by the TRS. The *z*-component of the spin polarization is negligible compared with the *x*- and *y*-components. Most interestingly, in Fig. 2(d) we plot the spin orientation θ, the angle between the spin and the +$k_x$ axis, as a function of $k_z$. It clearly shows a linear dependence following closely with Eq. (3b) as theoretically predicted from the first term of Eq. (2). This indicates that the SD in an elemental semiconductor locally breaks the inversion symmetry, thus generating a $k_z$-dependent local field pattern as depicted in Fig. 1(d). Therefore, generally in elemental semiconductors, a SD generates a SOC effect as described by the first term in Eq. (2).

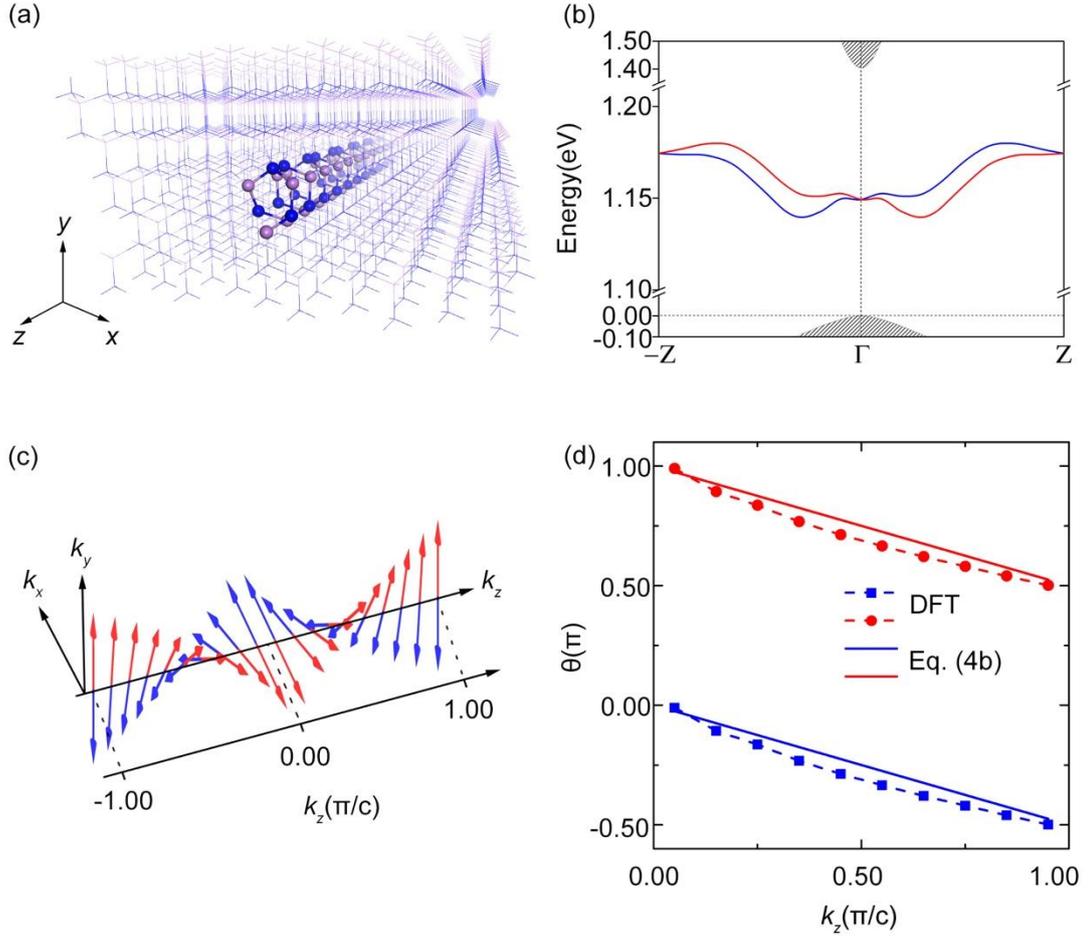

**Fig. 3.** (a) The atomic structure of SD in bulk GaAs. (b) Band structure of SD in GaAs with SOC effect. The Fermi level is set to zero. The red and blue lines represent SD-SOC bands of different spin projections. (c) The spin textures of SD-SOC bands obtained from DFT calculations. The red and blue arrows shows orientations of two spin projections. (d) Data points (dots) show θ, the angle between the spin and the $+k_x$ axis, as a function of $k_z$ obtained from DFT calculations. The lines are drawn according to Eq. (4b).

Next, we calculate the electronic structure of SD of GaAs (Fig. 3). As shown in Fig. 3(b), the SD-induced SOC states again resides inside the band gap. The calculated spin texture of GaAs is shown in Fig. 3(c). Again the z-component of the spin polarization is negligible compared with the x- and y-components. Interestingly, the spin texture is different from that of Ge, with spin orientations phase shifted by π/2. As plotted in Fig.

3(d), θ shows a linear dependence on $k_z$, closely following Eq. (4b), as theoretically predicted from the second term of Eq. (2). This result indicates that the SD in a compound semiconductor with large ionicity generates a $k_z$-dependent local field pattern as depicted in Fig. 1(e). Therefore, generally in compound semiconductors with large ionicity a SD generates a SOC effect as described by the second term in Eq. (2).

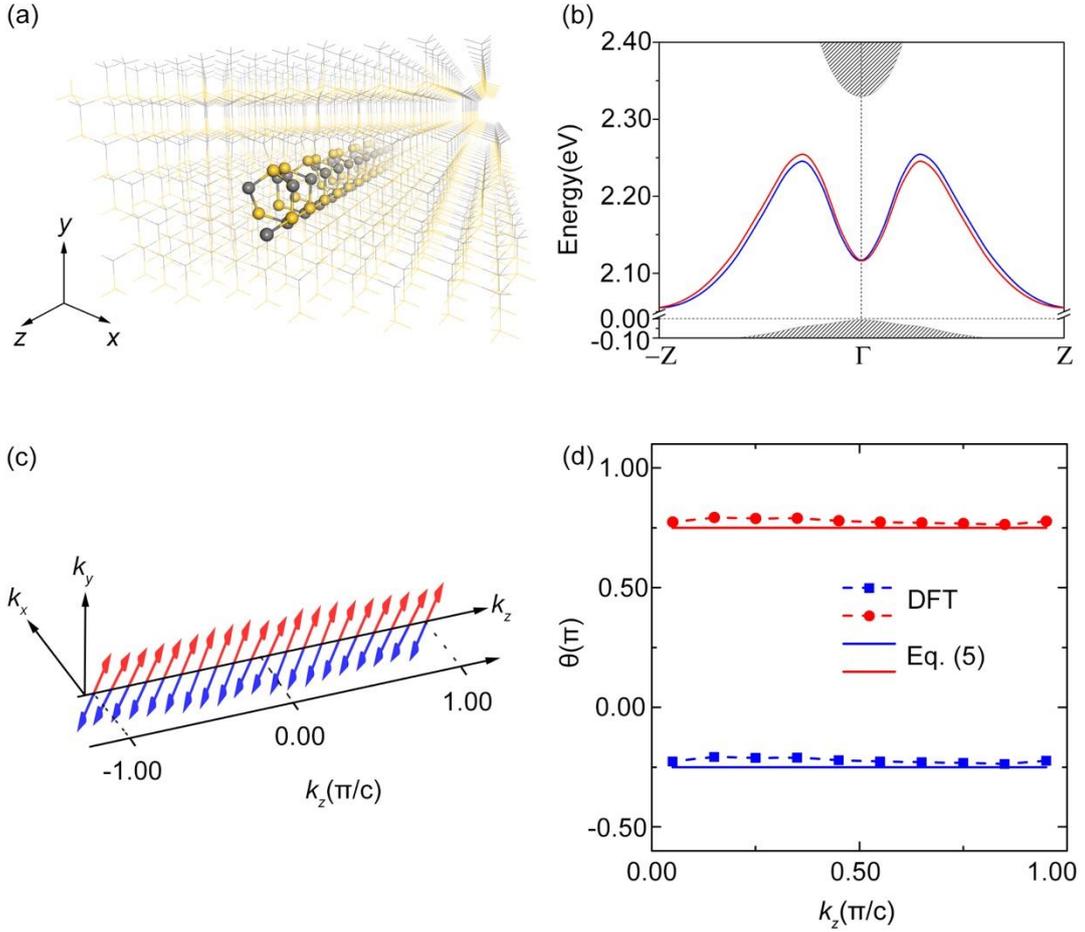

**Fig. 4.** (a) The atomic structure of SD in bulk SiC. (b) Band structure of SD in SiC with SOC effect. The Fermi level is set to zero. The red and blue lines represent SD-SOC bands of different spin projections. (c) The spin textures of SD-SOC bands obtained from DFT calculations. The red and blue arrows shows orientations of two spin projections. (d) Data points (dots) show θ, the angle between the spin and $+k_x$ axis, as a function of $k_z$ obtained from DFT calculations. The lines are drawn according to Eq. (5).

The calculations for Ge and GaAs can be considered to represent two extreme cases of an elemental semiconductor, with zero iconicity, and a compound semiconductor with large iconicity, respectively. Very interestingly, their spin textures turn out to correspond to the first and second term in Eq. (2) respectively. To explore this point farther, we have chosen SDs in SiC as an intermediate case representing a system with medium iconicity (Fig. 4). The SD in SiC has a similar structure (Fig. 4(a)) as in GaAs, as shown by experiments [22-23]. Fig. 4(b) shows the band structure of a SiC SD, with the SD-SOC states lying again inside the bandgap. A pleasant surprise is that the spin texture of SD in SiC, as shown in Fig. 4(c), is almost exactly the same as theoretically predicted in Eq. 5 and Fig. 1(f). The spin no longer rotates in the *xy*-plane, but is fixed in the [110] direction. Therefore, generally in compound semiconductors with medium iconicity a SD generates a SOC effect as described by both first and second terms in Eq. (2). It generates an ideal spin texture with spin conservation, which is predicted to be robust against all forms of spin-independent scattering [12,19,24]. Thus, the screw SD in those compound semiconductors with medium iconicity, like SiC, may be used effectively for suppressing spin relaxation in spintronics devices.

In all the above calculations, we have adopted the typical structures of SDs in semiconductors as discussed in the literature [20,22-23]. We have also calculated the stability of different SD structures and their corresponding SOC effects, which further confirmed the general applicability of our results and conclusions. For example, our first-principles calculations show that the SD is stable as it is in Si(Ge) and SiC without or with strain (~5% compression), respectively, but unstable in GaAs because of the formation of two wrong Ga-Ga and As-As bonds in the dislocation core. It is known that hydrogenation provides an effective means to stabilize the Ga-Ga bond in GaAs [25], so we have checked effects of strain and hydrogenation on the 1D SOC in SiC and GaAs, respectively. We found essentially the same results as before comparing Fig. S5 and S6 with Fig. 3 and 4, respectively [18]. Moreover, a SD is known to sometimes dissociate into two partial dislocations. For the case of GaAs, the dissociation reaction can be written as: 1/2 [1-10]→1/6 [1-21]+1/6 [2-11] [26]. We note that the rotational symmetry is changed from the two-fold rotation in the original SD to the six-fold rotation in the partial. In Fig. S7, we show the results for a pair of 90º partial dislocations in GaAs [18]. Again the results are qualitatively the same as in Fig. 3. However, there are some interesting quantitative differences. First, the two wrong bonds are split with the As-As and Ga-Ga bond in the left and right partials, respectively, as shown in Fig. S7a. The 1D SOC states in the bulk gap (Fig. S7b) come completely from the left core with the As-As bond (Fig. S7a), while the defect states coming from the right core are moved deep into the valence band due to hydrogenation. Secondly, the spin texture in Fig. S7a shows that the spin rotates only in the range of (0, $\pi/6$) due to the six-fold rotational symmetry in the partial dislocation, instead of (0, $\pi/2$) due to the two-fold rotational symmetry in the original SD (Fig. 3(c)). This is completely consistent with our theory.

Because the unique spin texture of the 1D SD-SOC is induced by the intrinsic helical symmetry of the SD, all the properties we reveal here will be ubiquitous in semiconductors with SDs. To supplement our first-principles calculations, we have also constructed a general (2×2) TB model Hamiltonian [See Equation (S6) [18]] to extract the quantitative strength of SOC in different materials (Table S3) by fitting the TB bands to the first-principles bands (Fig. S4). We found that for Ge, $\lambda_e = 0.0014$ is more than one order of magnitude larger than $\lambda_c = 0.0001$, indicating the first term in Eq. (2) dominates the SD-SOC in elemental semiconductors. In contrast, for GaAs $\lambda_e = 0.0012$ is almost one order of magnitude smaller than $\lambda_c = 0.0092$, so that the second term in Eq. (2) dominates the SD-SOC in semiconductors with large ionicity. For SiC, $\lambda_e = 0.0017$ is almost the same as $\lambda_c = 0.0018$, and the two terms in Eq. (2) contribute equally in semiconductors with medium ionicity. These results affirm again that the detailed spin texture can be tuned by the degree of iconicity with the same geometry of SDs, or by different geometries of SDs. The latter deserves further investigation. Thus, effectively a SD can act as a 'SOC torque' to generate and conduct highly coherent spin currents, and SDs are therefore useful for spintronics device applications [12]. Especially, the SD in SiC is predicted to afford the most attractive testbed for future experiments, with an ideal form of SOC. The perspective of maximizing the strength of SOC in SDs [27] is also very appealing.

**Acknowledgements**

We thank S. B. Zhang, S. -H. Wei, and X. W. Zhang for help discussions. L.H. and B.H. acknowledge the support from Science Challenge Project (No. TZ2016003), China Postdoctoral Science Foundation (No. 2017M610754), NSFC (Grant No. 11574024 and No. 11704021) and NSAF (No. U1530401). H.H. W.J., X.N. Y.Z and F. L. acknowledges the support from US-DOE (Grant No. DE-FG02-04ER46148). The work of V.Z. & M.G.L. was supported in part by NSF (Grant No. DMR-9632527 and DMR-9304912), DOE (Grant No. DE-FG02-00ER45816) and the Alexander von Humboldt Foundation. Computations were performed at Tianhe2-JK at CSRC.